\documentclass[referee]{raa}
\usepackage{graphicx,times}             
\input{psfig.sty} 

\begin{document}
 
\title{Photometric study of an Eclipsing Binary in Praesepe}

\author{Devarapalli Shanti Priya$^{1}$, Kandulapati Sriram$^{2}$
           \and 
            Pasagada Vivekananda Rao$^{1}$ }

\institute{ 1. Department of Astronomy, Osmania University,
              Hyderabad 500 007, India\\
                {\it astroshanti@gmail.com}\\
              2. Korea Astronomy and Space Science Institute, Daejeon, South Korea\\ 
           }

\abstract{We present CCD photometric observations of an eclipsing binary in the direction of the open cluster Praesepe using the 2 m telescope of IUCAA Girawali Observatory, India. Though the system was classified as an eclipsing binary by Pepper et al.(2008),detail investigations were lacking. The photometric solutions using the Wilson-Devinney code suggest that it is a W-type W UMa system and interestingly, the system parameters were similar to another contact binary system SW Lac. 
}

\authorrunning{Shanti, Sriram \& Rao }           
\titlerunning{W-type W UMa binary}

\maketitle

\section{Introduction}
Among eclipsing binaries, short period low temperature ($\le$ 1 d) contact binaries are important astrophysical systems (Yakut \& Eggelton 2005). They are often found in fields as well as in galactic and globular clusters with relatively high frequencies (Ruscinski 1997, 1998 \& 2000). The contact binaries in clusters are of great importance as their evolution can be understood on the basis of host cluster age; however no concrete correlation is yet revealed. 
The theoretical understanding of their origin, evolution, thermal equilibrium and mass exchange are least understood; however detailed modeling suggest that they coalesce at their final stage (Stepien et al. 2006). On the other hand, for few contact binaries the observational properties of the binary components are known accurately (eg. 52 W-type and 60 A-type W UMa systems see table 1 and 2 of Gazeas \& Stepien 2008). 

Pepper et al. (2008) have obtained light curves for 66,638 stars in NGC 2632 (Praesepe) out of which 208 were variables and among them $\sim$100 are categorized as eclipsing binaries. Four eclipsing binaries among them have been classified to be W UMa type systems (TX Cnc, EF Cnc, GW Cnc, EH Cnc) and the observational properties of remaining eclipsing binaries are yet to be studied whose periods falls in the range from 0.26 -- 11.03 days.   
One of the variables (KELT ID KP101231) was selected for photometric studies. The J, H, K magnitudes of the variable are 9.794, 9.391 and 9.272 respectively and R$_{k}$ is 10.987. The period of the selected variable was found to be 0.2910 d with an amplitude of $\sim$0.5.
Our work draws motivation from the fact that the observational properties of very few contact binary systems of such short period are known and hence 
we present the detailed V passband photometric solutions for this variable using Wilson-Devinney code (2003 version).

\section{Observations and Data Reduction}

 The V band CCD photometric observations of the variable were carried out using the 2-m IGO telescope, India. The observations were performed during two nights on February 26 and 27, 2012.  The IUCAA Faint Object Spectrograph Camera (IFOSC) was used, which is equipped with EEV 2 K $\times$ 2 K thinned, back-illuminated CCD with 13.5$\mu$m pixels. The CCD used for imaging provides an effective field of view of $\sim$10$\times$10 arcmin$^{2}$ on the sky corresponding to a plate scale of 0.3 arcsec per pixel. A total of 289 frames in the V band were obtained with an exposure time of 10 -- 35 s for a good photometric accuracy and the field was centered at $\alpha$$_{J2000}$=08$^{h}$ 57$^{m}$ 9.71$^{s}$,  $\delta$$_{J2000}$=18$^{o}$ 56$^{'}$ 44.12$^{''}$.
 
We have performed aperture photometry with the {\it apphot} package available in IRAF software. The Figure 1 shows the positions of the variable (V1), comparison (C1) and check (C2) stars. It was found that the comparison and check stars were relatively constant in brightness. From the light curve (variable - comparison) obtained, it was found that the shape and depths are similar to the published light curve by Pepper et al. (2008). The Figure 2 shows the light curve $\delta$m(Comparison-Check) vs HJD, which is constant in brightness.
\section{Photometric solutions}
The visual inspection 
of the variable resemble W UMa type binary nature
and hence mode 3 was finally adopted after noticing that mode 2 is converging to the mode 3.
The photometric solutions were obtained by using Wilson-Devinney
program with an option of non-linear limb darkening via a square root
law along with many other features (van Hamme \& Wilson 2003). We too determined the 
period of the variable to be P = 0.2910 d using the Lomb-Scargle periodogram (Scarlge 1982; Zechmeister \&  K¨¹rster 2009), which is similar to the value obtained by Pepper et al. (2008).
 We have adopted the procedure of fixing the temperature, limb darkening coefficients, x$_{h}$ \& x$_{c}$,  gravity-darkening coefficients g$_{h}$ \& g$_{c}$ and albedos, A$_{h}$ \& A$_{c}$ for the components as described in our earlier papers (Rukmini
\& Vivekananda Rao 2002, Rukmini et al. 2005, Sriram et al. 2009, 2010; Shanti et al. 2011 and Ravi kiron et al. 2011, 2012).
The mass ratio is an important parameter for a binary system and is a pivotal 
parameter to know the evolution of an contact binary system. To constrain the mass ratio parameter, initially, 
mean effective temperature of cool star T$_{c}$, the dimensionless surface potentials $\Omega_{h}$=$\Omega_{c}$, and
the monochromatic luminosity L$_{h}$(V) along with orbital inclination {\it i} are adopted as adjustable
parameters. To determine the accurate value of mass ratio, this parameter is also taken as an adjustable one along with the others, till a convergent solution is obtained. The Figure 3 shows the weighted sum of square of residuals ($\Sigma W(O-C)^{2}$)) for different assumed mass ratio values and the best value was found to be 1.29 $\pm$ 0.01 for variable V1. The results of the computed solutions 
are shown in Table 1. To obtain the theoretical light curves, we have used the LC program incorporating the final parameters resulted from the DC program. The observed and theoretical light curves for the variable are shown in Figure 4.

\begin{figure}
\includegraphics[height=10cm,width=15cm, angle=0]{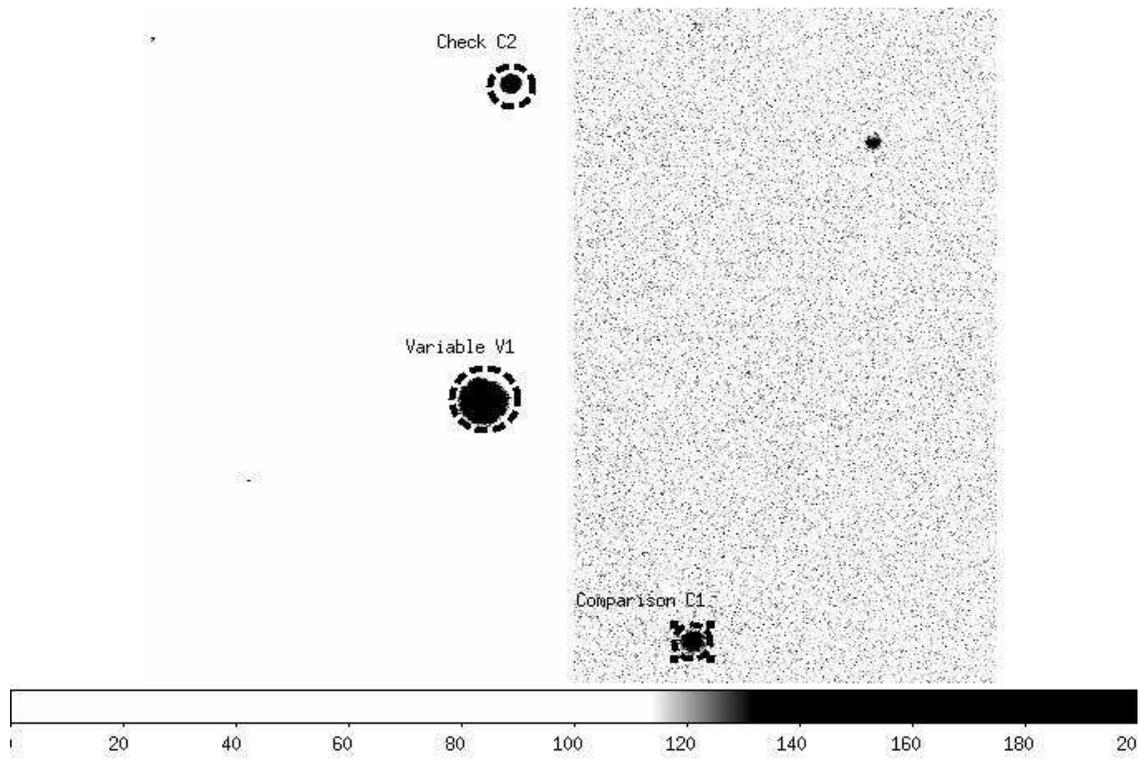} \\

\caption{$\sim$10$\times$10 arc min image of the field of NGC 2632.The variable, comparison and the check star are marked. \label{Fig1}
}
\end{figure}


\begin{figure}
\includegraphics[height=10cm,width=15cm, angle=0]{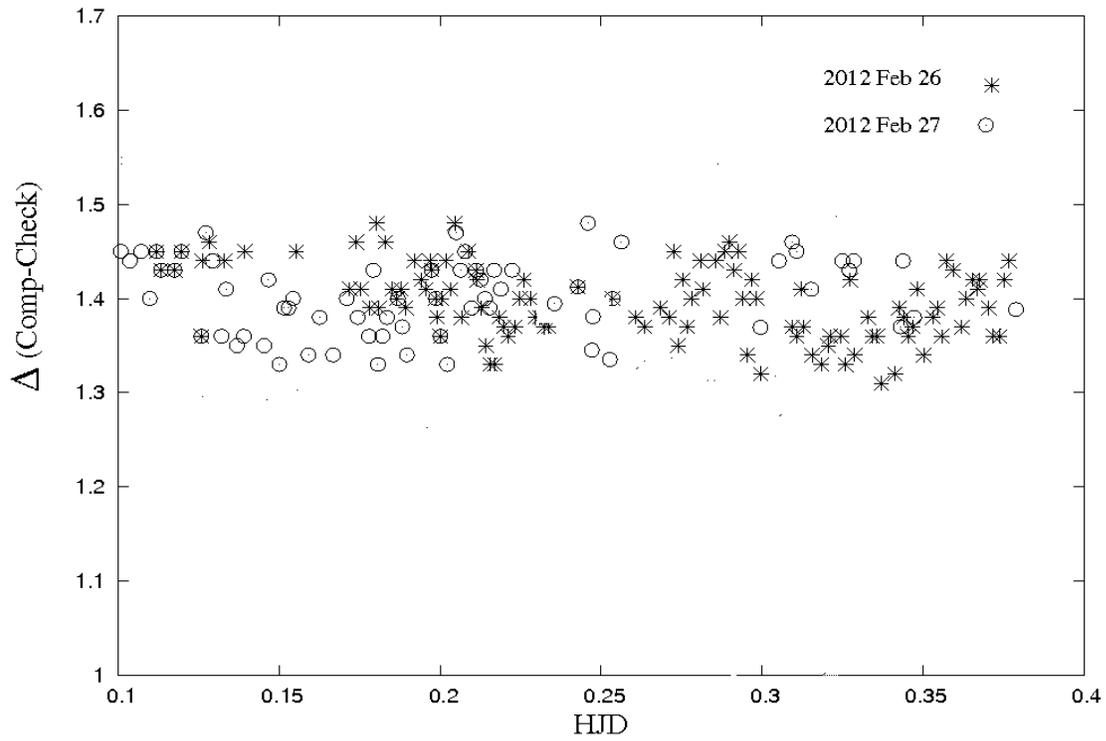} \\

\caption{The figure shows the magnitude difference between comparison and check star versus HJD(2455984+ \& 2455985+) for observations of two nights respectively. \label{Fig2}
}
\end{figure}


\begin{figure}
\includegraphics[height=15cm,width=10cm, angle=-90]{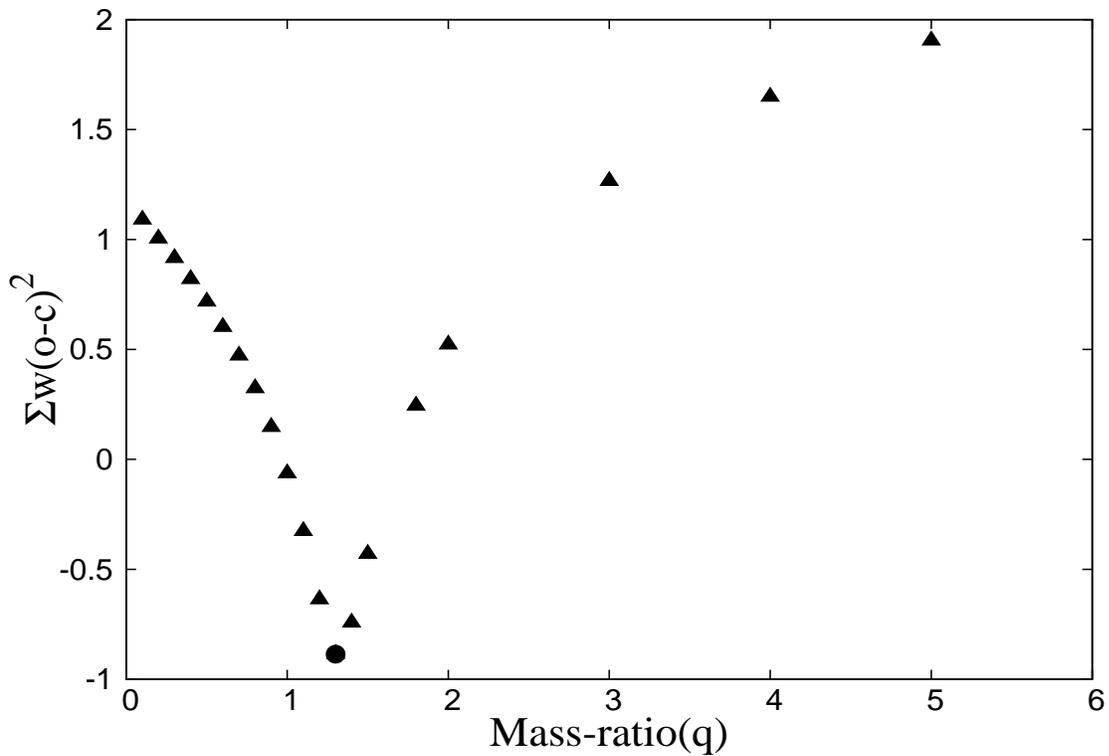} \\

\caption{The figure shows the q versus $\Sigma$w(o-c)$^{2}$ for the variable V1. \label{Fig3} 
}
\end{figure}

\begin{figure}
\includegraphics[height=15cm,width=10cm, angle=0]{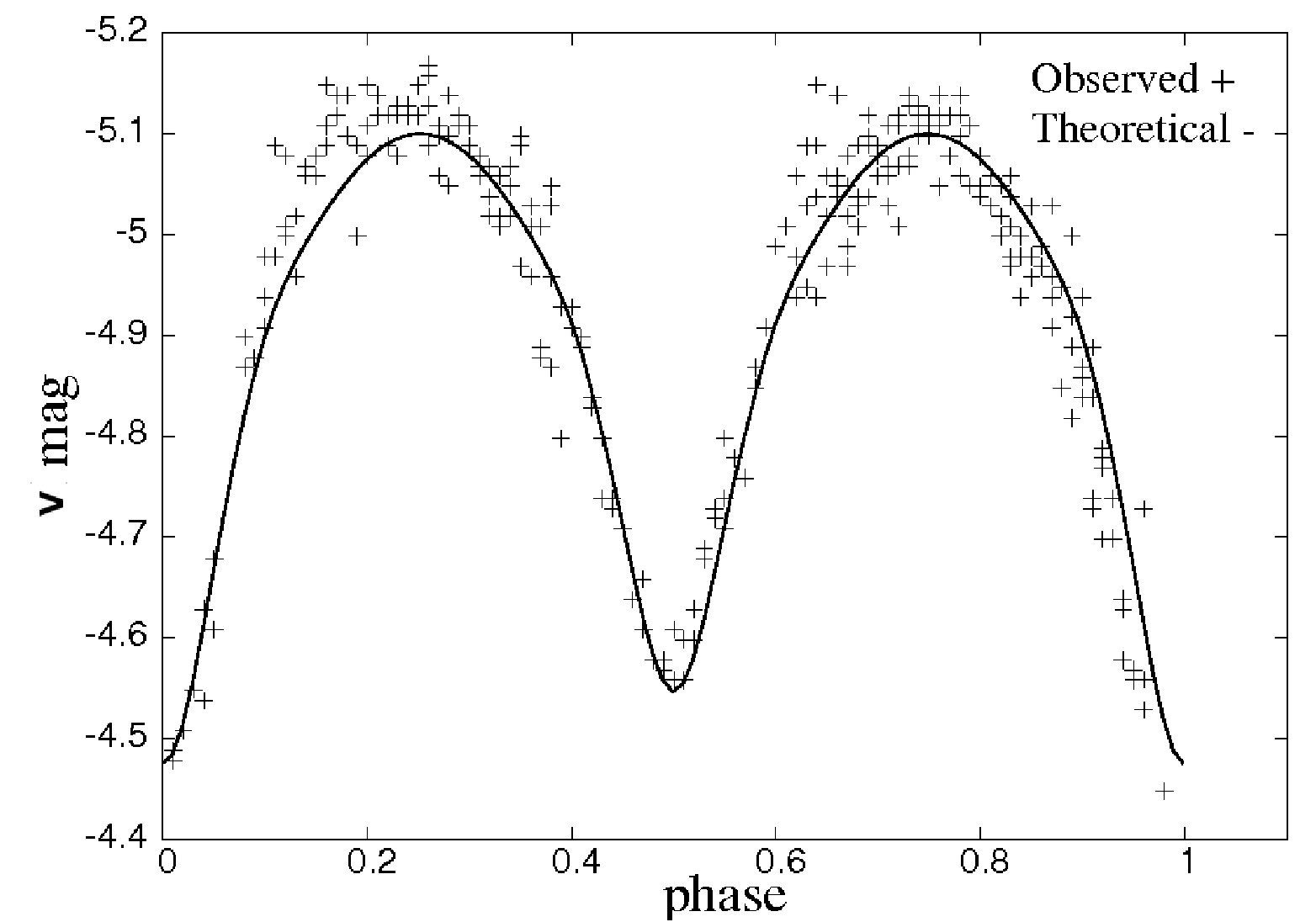} \\

\caption{The figure shows the best fit to the V passband light curve for the variable V1. \label{Fig4}
}
\end{figure}

\clearpage

{\begin{table*}
\begin{minipage}[t]{\columnwidth}
\caption[solutions]{The Photometric elements obtained for Variable(V1) by using W-D method.}
\label{Table 1}
\renewcommand{\footnoterule}{}
\centering
\begin{tabular}{ccccccc}
\hline
\hline
Element & & V1\\ \\
\hline
\hline
Period (days)&&0.2910\\
$T_{e,h}$ K & & 5477 \\ 
$T_{e,c}$\,K& & 5309$\pm$6 \\
q & & 1.29$\pm$0.01 \\ 
i$^{o}$& & 76.22$\pm$0.10 \\ 
$\Omega$ &  &4.1580$\pm$0.0188 \\
fill-out factor &     &0.1104 \\
$r_{h}$ & pole & 0.3408$\pm$0.0024 \\
        & side & 0.3579$\pm$0.0029 \\
        & back & 0.3929$\pm$0.0046 \\

$r_{c}$ & pole & 0.3835$\pm$0.0022 \\
 	& side & 0.4054$\pm$0.0028 \\
	& back & 0.4382$\pm$0.0041 \\

$L_{h}$& & 0.5668$\pm$0.2718 \\

$L_{c}$ & & 0.6142 \\ 


$x_{h}$ & & 0.60$\pm$0.05 \\ 

$x_{c}$ & & 0.60$\pm$0.05 \\
$\Sigma$w(o-c)$^{2}$ & &0.01954 \\
Spectral type & &G6-K0  \\

A$_{h}$& & 0.5 \\ 
A$_{c}$& & 0.5 \\ 
\hline
\hline
\end{tabular} 
\end{minipage}
\end{table*}}

\clearpage

\section{Discussion and Result}

The study of contact binary systems is important in order to understand the physics of their formation and evolution, it also gives some information of the parent cluster if it happens to be a member. However the selected variable was not found to be the member of the cluster NGC 2632 (Pepper et al. 2008) but the present study is a preliminary investigation of a short period binary,and the number of such short period contact binary systems are less studied. The period of the system is around  $\sim$0$^{d}$.2910 and the J-H=0.40 value corresponds to G6-K0 
spectral type (based on Allen's table). The best fit solution revealed that the temperature difference between the components is $\sim$168 K suggesting that they are at good degree of thermal contact. 
The best combination of q and i is 1.29 \& $\sim $76.22$^{o}$ respectively. The fill-out factor is 0.1104 i.e. the two components of binary systems are at $\sim$ 11\% geometrical contact.  
The solutions indicate that the V1 is a W-type W UMa type variable. 
Gazeas (2009) obtained three dimensional correlation using primary's mass, period and mass ratios which are stated below:\\
 log $M_{1}$ = 0.725log P - 0.076log q + 0.365 \\
 log $M_{2}$ = 0.725log P + 0.924log q + 0.365 \\
 log $R_{1}$ = 0.930log P - 0.141log q + 0.434 \\
 log $R_{2}$ = 0.930log P + 0.287log q + 0.434 \\
 log $L_{1}$ = 2.531log P - 0.512log q + 1.102 \\
 log $L_{2}$ = 2.531log P + 0.352log q + 1.102 \\
Substituting the mass ratio from the best fit solutions, we derived the following M$_{1}$= 0.9288 M$_{\odot}$, M${_2}$= 1.1982 M$_{\odot}$, R${_1}$= 0.5752 R$_{\odot}$, R${_2}$ =0.5129 R$_{\odot}$
L${_1}$=  0.4881 L$_{\odot}$, L${_2}$ = 0.6082 L $_{\odot}$. 
The solutions of the variable (V1) were found to be similar to another contact binary system properties, SW Lac (P = 0.32 d, T1 = 5800 K, T2 = 5515 K, q = 1.270, and i = 80, f=30\%  Gazeas et al. (2005). It can be observed that mass ratio, inclination and radius parameters are closely matching.


\begin{acknowledgements}
   The authors acknowledge Helmut Abt for his appreciative and encouraging comments.  
\end{acknowledgements}

\end{document}